\documentclass[9pt,technote]{IEEEtran}
\usepackage{graphics}

\begin{document}

\title{Geometric symmetry in the quadratic Fisher discriminant
operating on image pixels}

\author{Robert~S.~Caprari%
\thanks{Manuscript received ????????; revised ????????.}%
\thanks{R.~Caprari is with the Defence Science and Technology Organisation (DSTO),
PO Box 1500, Edinburgh SA 5111, Australia. (email:
robert.caprari@dsto.defence.gov.au)}}

\markboth{IEEE Transactions on Information Theory,~Vol.~??,
No.~??,~????????~????}{Caprari: Geometric symmetry in the
quadratic Fisher discriminant operating on image pixels}

\pubid{0000--0000/00\$00.00~\copyright~???? IEEE}

\maketitle

\begin{abstract}
This article examines the design of Quadratic Fisher Discriminants
(QFDs) that operate directly on image pixels, when image ensembles
are taken to comprise all rotated and reflected versions of
distinct sample images. A procedure based on group theory is
devised to identify and discard QFD coefficients made redundant by
symmetry, for arbitrary sampling lattices. This procedure
introduces the concept of a degeneracy matrix. Tensor
representations are established for the square lattice point group
(8-fold symmetry) and hexagonal lattice point group (12-fold
symmetry). The analysis is largely applicable to the
symmetrisation of any quadratic filter, and generalises to higher
order polynomial (Volterra) filters. Experiments on square lattice
sampled synthetic aperture radar (SAR) imagery verify that
symmetrisation of QFDs can improve their generalisation and
discrimination ability.
\end{abstract}


\IEEEpeerreviewmaketitle

\section{Introduction}
\label{sec: intro}

\PARstart{I}{n} image target detection, one often desires to
detect all geometric symmetry transformed versions of targets, for
two reasons. One reason is that for any given target pattern, all
symmetry transformed versions also are valid target patterns, to
at least a good approximation. The other reason is that so few
separate target patterns are available for detector training,
relative to the number of detector coefficients, that the detector
has the capacity to learn the peculiarities of each individual
pattern, rather than being forced to learn the universal
properties of the target pattern class. Symmetry transformation of
available target patterns creates additional patterns, which even
if not valid target patterns, at least share many of the universal
properties of genuine target patterns. Training the detector on
both actual and transformed target patterns discourages
overlearning of individual pattern peculiarities, and encourages
learning of universal pattern properties.

There are four distinct classes of geometric point
transformations\footnote{ Point transformations are geometric
transformations that leave at least one point invariant.}:
rotation; reflection; shear; and dilation. Rotation is a
sense-preserving rigid-body transformation. Reflection is a
sense-reversing rigid-body transformation. Shear is an
angle-modifying, area-preserving transformation. Dilation is an
angle-preserving, area-modifying transformation. Usually only
rotation operations are considered in image target detection
\cite{Hsu82}--\cite{Fukumi97}. Reflection operations also are
valid where targets have approximate mirror symmetry, or where the
second reason in the previous paragraph is the motivating factor
for considering symmetry. Dilation and shear have the fundamental
theoretical drawback that no matter how symmetrical the detector
support, there is always a flux of image content into or out of
the support under these transformations, so that the detector
support is not transformed onto itself. Dilations have the
additional theoretical drawback for sampled (discrete) images that
all strict dilations map non-lattice points onto lattice points,
and all strict contractions map lattice points onto non-lattice
points; once again, the discrete-space detector support is not
mapped onto itself. These properties of dilation and shear
preclude their inclusion in group theoretical methods of
accounting for geometric symmetry\footnote{An example of the
problem is that a dilation followed by its reverse contraction is
not an identity transformation, because the original dilation
ejected the periphery of the support, and the following
contraction leaves a void in the same periphery, even though the
interior of the support has been perfectly restored.} in pattern
recognition, as developed here and
elsewhere~\cite{Lenz89}--\cite{Liu04}. Group theoretical
considerations make rotation and reflection the only allowable
geometric symmetries in this analysis. Even then, the detector
support requires at least the same rotation and mirror symmetry,
and orientation\footnote{A polygon and lattice of the same
symmetry have the same orientation if their mirror lines
coincide.}, as the sampling lattice, for the discrete space
support to map onto itself, as it must. If one treats images as
being spatially continuous and detector supports infinitely large,
then dilation and shear are permissible symmetry
operations~\cite{Leen95}.

\pubidadjcol

This article presents the group theory based geometric symmetry
analysis of the Quadratic Fisher Discriminant (QFD) operating on
image pixels. Encapsulated in the analysis is a method of
symmetrising any quadratic detector, not just the QFD\@. The
mathematical formalism generalises in a straightforward manner to
polynomial detectors of arbitrary degree (Volterra filters).
Symmetrisation of the QFD, or indeed any polynomial filter, is a
means of reducing the number of detector coefficients by
identifying and discarding redundant coefficients, without
introducing approximations. In contrast, most approaches to
reducing polynomial filter complexity approximate the desired
filter by a simpler version~\cite{Nowak96}.

A synopsis of this article is as follows. Section~\ref{sec: square
lattice} introduces notation and establishes the square lattice
symmetry group. Section~\ref{sec: invariants} derives the
consequences for the QFD of sampling lattice symmetry.
Section~\ref{sec: D matrix} solves for the symmetrised QFD, using
a procedure that is suitable for arbitrary quadratic filters, and
that easily extends to higher degree polynomial filters.
Section~\ref{sec: hexagonal lattice} establishes the hexagonal
lattice symmetry group, and notes the detail changes in the
preceding analysis if images are sampled on an hexagonal lattice
instead of a square lattice. Section~\ref{sec: experiment}
experimentally demonstrates the symmetrisation of a QFD for target
detection in synthetic aperture radar (SAR) images sampled on a
square lattice.

\section{Square lattice symmetry}
\label{sec: square lattice}

\begin{figure}[t!]
\begin{center}
\includegraphics{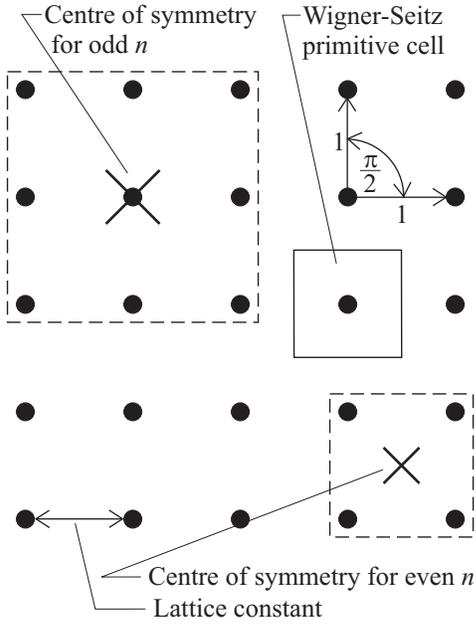}
\end{center}
\caption{(Refer to Section~\ref{sec: square lattice}) A section of
square lattice. Nonequivalent points marked $\times$ have
identical symmetry. Detector supports of odd linear dimension
centre on lattice points. Detector supports of even linear
dimension centre on interstices.} \label{fig: sq_net}
\end{figure}

As shown in Figure~\ref{fig: sq_net}, the square lattice is
generated by two equal length basis vectors subtending angle
$\pi/2$. To exploit the symmetry of the sampling lattice, the
detector support must be a polygon with the same symmetry as the
lattice, except for translational symmetry. Conventionally, a
square lattice is taken to use a square detector support of
dimension $n\!\times\!n$ pixels, with total pixel count
\begin{equation}
\label{sq: N def} N \equiv n^{2} \ ,
\end{equation}
and with lattice basis vectors as perpendicular bisectors of its
sides. Although not considered here, another (unusual) possibility
is a square detector support centred on a lattice point and
oriented so that the lattice basis vectors point at its vertices
(i.e. a diamond shape).

The set of symmetry operations that rigidly transform a lattice
onto itself while leaving one lattice point fixed, form the point
group of the lattice\footnote{The space group of a lattice
comprises point symmetries plus translational symmetries.}. The
fixed lattice point is the centre of symmetry. Point symmetries
are restricted to rotations and reflections, since dilations and
shears are prohibited for the reasons given in Section~\ref{sec:
intro}\@. For a square lattice, the point group symmetry
operations with centre displaced by one half lattice spacing
horizontally and vertically from a lattice point, also map the
square lattice onto itself, albeit without leaving any lattice
point fixed (Figure~\ref{fig: sq_net}). If the detector support
has odd linear dimension (odd~$n$) then it is centred on a lattice
point, whereas if the detector support has even linear dimension
(even~$n$) then it is centred on an interstitial site. Although
the two centres are not equivalent\footnote{Nonequivalent points
are separated by a nonintegral combination of lattice vectors.},
the square lattice has identical symmetry about both centres, and
that symmetry is the point symmetry.

A symmetry group {\boldmath$T$} having $s$ elements $T_{i}$, that
is,
\begin{equation}
\mbox{\boldmath $T$} \equiv \{ T_{1},\ldots,T_{s} \} \ ,
\end{equation}
represents symmetry of order~$s$ (i.e.\ $s$-fold symmetry). Square
lattice point symmetry is of order
\begin{equation}
s=8 \hspace{2.0em} \mbox{(square lattice)} \ .
\end{equation}
Table~\ref{tab:sq_point_group} lists the 8 elements of the square
lattice point group---being the dihedral group
$D_{4}$\footnote{Dihedral group $D_{n}$ is the group of symmetry
transformations of an $n$-sided regular polygon, being an $n$-fold
rotation axis and $n$ mirror lines.}---in the first column. The
second column of Table~\ref{tab:sq_point_group} describes the
geometric operations corresponding to the point symmetries.
Symmetries $T_{1}$ to $T_{s}$ satisfy the four mandatory group
properties: closure; presence of identity element; presence of all
inverse elements; and associativity. The point group is
non-Abelian, because in general symmetry operations do not commute
(e.g.\ for the square lattice point group
$T_{5}\,T_{6}=T_{4}\neq\,$$T_{2}=T_{6}\,T_{5}$).

\begin{table*}[t!]
\begin{center}
{\footnotesize
\begin{tabular}{|l|l|l|l|}
\hline \rule{0em}{2.5ex}

Symmetry: & Description of symmetry & Factorisation &
Rank 2 tensor representation:   \\
$T_{m}$   & operation               &               &
$t^{(2)}_{m}(\{ij\};\{i'j'\})$ \\[0.5ex]

\hline \hline \rule{0em}{2.5ex}

$T_{1}$ & Identity or unit operation.  & $T_{5}^{\,2}$      &
$\delta(i,i').$ \\
        &                              &                  &
$\delta(j,j')$  \\[0.5ex]

\hline \rule{0em}{2.5ex}

$T_{2}$ & Anticlockwise rotation by    & $T_{2}$          &
$\delta(i,n-j'+1).$ \\
        & 1/4 turn.                    &                  &
$\delta(j,i')$ \\[0.5ex]

\hline \rule{0em}{2.5ex}

$T_{3}$ & Rotation by 1/2 turn, being  & $T_{2}^{\,2}$      &
$\delta(i,n-i'+1).$ \\
        & inversion through centre.    &                  &
$\delta(j,n-j'+1)$ \\[0.5ex]

\hline \rule{0em}{2.5ex}

$T_{4}$ & Clockwise rotation by 1/4   & $T_{2}^{\,3}$      &
$\delta(i,j').$ \\
        & turn.                        &                  &
$\delta(j,n-i'+1)$ \\[0.5ex]

\hline \rule{0em}{2.5ex}

$T_{5}$ & Reflection in the $x$-axis.  & $T_{5}$          &
$\delta(i,n-i'+1).$ \\
        &                              &                  &
$\delta(j,j')$ \\[0.5ex]

\hline \rule{0em}{2.5ex}

$T_{6}$ & Reflection in line 1/8 turn  & $T_{2}\,T_{5}$     &
$\delta(i,n-j'+1).$ \\
        & anticlockwise from $x$-axis. &                  &
$\delta(j,n-i'+1)$ \\[0.5ex]

\hline \rule{0em}{2.5ex}

$T_{7}$ & Reflection in the $y$-axis.  & $T_{2}^{\,2}\,T_{5}$ &
$\delta(i,i').$ \\
        &                              &                  &
$\delta(j,n-j'+1)$ \\[0.5ex]

\hline \rule{0em}{2.5ex}

$T_{8}$ & Reflection in line 1/8 turn  & $T_{2}^{\,3}\,T_{5}$ &
$\delta(i,j').$ \\
        & clockwise from $x$-axis.     &                  &
$\delta(j,i')$ \\[0.5ex]

\hline
\end{tabular}}
\end{center}
\caption{(Refer to Section~\ref{sec: square lattice}) Square
lattice point group symmetry operations and their rank 2 tensor
representation; the rank 4 tensor representation is derived from
the rank 2 tensor representation according to (\ref{t(4) from
t(2)}). The detector support is square with sides of length
$n$-pixels. Orientation of the lattice is such that a basis vector
aligns with the $x$-axis, as in Figure~\ref{fig: sq_net}.}
\label{tab:sq_point_group}
\end{table*}

Subgroups of {\boldmath $T$} represent image ensembles with
intermediate geometric symmetry. For square lattice sampling: $\{
T_{1}, T_{2}, T_{3}, T_{4}\}$ represents a 4-fold rotation axis;
$\{ T_{1}, T_{3}, T_{5}, T_{7}\}$ and $\{ T_{1}, T_{3}, T_{6},
T_{8}\}$ represent 4-fold rectangular symmetry; $\{ T_{1},
T_{3}\}$ represents a 2-fold rotation axis; $\{ T_{1}, T_{5}\}$,
$\{ T_{1}, T_{6}\}$, $\{ T_{1}, T_{7}\}$ and $\{ T_{1}, T_{8}\}$
represent 2-fold reflection symmetry (one mirror line); and
$\{T_{1}\}$ represents the complete absence of geometric symmetry.
In the context of airborne surveillance, imaging in the nadir
direction has the full symmetry of group {\boldmath $T$}, but
side-looking SAR only has a mirror line along the range direction
($y$-axis), corresponding to subgroup $\{ T_{1}, T_{7}\}$.

Introduce the 2-component index (2-index) $\{ij\}$ to convert the
two indices of detector pixels (row $i$ and column $j$) into a
single index,
\begin{equation}
\label{2-component index def} \{ij\} \equiv i + (j-1)n \, : \ i,j
\!\in\! [1,n] \, , \ \{ij\} \!\in\! [1,N] \ ,
\end{equation}
thereby allowing pixels in the support to be column-ordered into
the dimension $N$ column vector {\boldmath $x$}$^{(1)}$ with
components $x^{(1)}(\{ij\})$, where the (1) superscript signifies
that {\boldmath $x$}$^{(1)}$ is inherently a rank 1 tensor (i.e.\
vector). It is possible to construct QFD theory using 2-indices
alone, at the expense of using rank 4 tensors. Alternatively,
combining pairs of 2-indices ($\{ij\}$ and $\{kl\}$) into
4-component indices ($\{ij,kl\}$) according to the prescription
\begin{eqnarray}
\label{4-component index def} \lefteqn{\{ij,kl\} \equiv \{ij\} +
(\{kl\}-1)N \, :} \nonumber \\
 & & \{ij\},\{kl\} \!\in\! [1,N] \, , \ \{ij,kl\} \!\in\!
 [1,N^{2}] \ ,
\end{eqnarray}
allows rank 4 tensors to be expressed as higher dimension rank 2
tensors (i.e.\ square matrices), as will be seen forthwith. As a
reminder that 4-indices are properly a pair of 2-indices, the
notation $\{ij,kl\}$ retains a comma between the first and second
2-indices. Pairwise products of pixels in the support are
collected into the dimension $N^{2}$ column vector {\boldmath
$x$}$^{(2)}$ with components
\begin{equation}
\label{x_2 components} x^{(2)}(\{ij,kl\}) \equiv x^{(1)}(\{ij\})
\, x^{(1)}(\{kl\}) \ ,
\end{equation}
where the (2) superscript signifies that {{\boldmath $x$}$^{(2)}$
is inherently a rank 2 tensor\footnote{$x^{(2)}$ is the outer
product of $x^{(1)}$ with itself.}.

Symmetry operation $T_{m}$ linearly transforms unprimed vectors
into primed vectors as follows:
\begin{equation}
\label{symmetry transformation algebra} \mbox{\boldmath
$x$}^{(i)\prime} = T^{(2i)}_{m} \, \mbox{\boldmath $x$}^{(i)} \ .
\end{equation}
$T^{(2)}_{m}$, with components $t^{(2)}_{m}(\{ij\};\{i'j'\})$, is
the rank 2 tensor representation of symmetry operation $T_{m}$.
$T_{m}^{(4)}$, with components
$t^{(4)}_{m}(\{ij,kl\};\{i'j',k'l'\})$, is the rank 4 tensor
representation of symmetry operation $T_{m}$. Use of 4-index
notation allows $T_{m}^{(4)}$ to be expressed in matrix form, but
this is merely a notational convenience. Lattice point symmetries
rearrange image pixels without altering pixel values (i.e. they
are permutation operations on the pixel values), implying
invariance of the Euclidean norm\footnote{This length property
establishes that tensors are defined in a Euclidean vector space,
so they are cartesian tensors with no distinction between
covariant and contravariant indices.}
\begin{equation}
\mbox{\boldmath $x$}^{(i)\prime\,\mathrm{T}} \, \mbox{\boldmath
$x$}^{(i)\prime} = \mbox{\boldmath $x$}^{(i)\,\mathrm{T}} \,
\mbox{\boldmath $x$}^{(i)} \ ,
\end{equation}
which ensures that all transformation matrices are orthogonal,
that is,\footnote{The transpose of a rank 4 tensor with four
1-component indices is defined as the swapping of the first and
second index pairs, with the index ordering in individual pairs
not changing (i.e.\ $t(i,j,k,l)$ becomes $t(k,l,i,j)$).}
\begin{equation}
\label{matrix orthogonality} T^{(2i)\,-1}_{m} =
T^{(2i)\,\mathrm{T}}_{m} \, , \ m=1,\dots,s \ .
\end{equation}
$T^{(2i)}_{m}$ have determinants 1 or $-1$\footnote{\label{fn:
determinants}It is not the case that all rotations have
determinant 1 and all reflections have determinant $-1$, because
the matrices transform pixel values and not pixel coordinates (the
latter being the common usage for transformation matrices).}. For
a square lattice, $T^{(2i)}_{1}$ has eigenvalue 1, $T^{(2i)}_{3}$
and $T^{(2i)}_{5-8}$ have eigenvalues 1 and $-1$, and
$T^{(2i)}_{2}$ and $T^{(2i)}_{4}$ have eigenvalues 1, $i$, $-1$
and $-i$. Expanding (\ref{symmetry transformation algebra}) in
terms of tensor components, and substituting (\ref{x_2
components}),
 connects components of the two representations by
\begin{equation}
\label{t(4) from t(2)} t^{(4)}_{m}(\{ij,kl\};\{i'j',k'l'\}) =
t^{(2)}_{m}(\{ij\};\{i'j'\}) \, t^{(2)}_{m}(\{kl\};\{k'l'\}) \ .
\end{equation}
Equation (\ref{t(4) from t(2)}) implies that the rank 4 tensor
components have permutation symmetry
\begin{equation}
t^{(4)}_{m}(\{ij,kl\};\{i'j',k'l'\}) =
t^{(4)}_{m}(\{kl,ij\};\{k'l',i'j'\}) \ .
\end{equation}

The rank 2 tensor representation of the square lattice point group
is listed in the fourth column of Table~\ref{tab:sq_point_group};
the rank 4 tensor representation is computed from the rank 2
tensor representation according to (\ref{t(4) from t(2)}). Each
$T^{(2i)}_{m}$ matrix has a single one entry in every row and
column. Computing the rank 2 tensor representation is efficiently
done as follows. Derive matrices $T^{(2)}_{2}$ and $T^{(2)}_{5}$
by inspection. Elements $T_{2}$ and $T_{5}$ generate group
{\boldmath $T$}, as indicated in the third column of
Table~\ref{tab:sq_point_group}, so the remaining elements of the
rank 2 representation of {\boldmath $T$} are computed by repeated
multiplication of matrices $T^{(2)}_{2}$ and
$T^{(2)}_{5}$\footnote{$T^{(2)}_{1}$, representing the identity
operation, is immediately set to the identity matrix, without
being explicitly generated as $T^{(2)\,2}_{5}$.}.

\section{Symmetry invariants}
\label{sec: invariants}

The underlying postulate of this analysis is that training
ensemble statistics (clutter and target separately) are invariant
with respect to all point symmetry transformations about the
centre of detector support. Such statistical symmetry arises if
every image in the actual ensemble contributes to a notional
ensemble all symmetry transformations of itself (with equal
weighting), and the notional ensemble is a more complete portrayal
of the totality of images than the actual ensemble. There is no
requirement for images in the ensembles to be individually
invariant with respect to any symmetry transformations.

The quadratic detector response is given by the scalar product
\begin{equation}
\label{detector operation} y(\mbox{\boldmath $x$}) =
\mbox{\boldmath $f$}^{\mathrm{T}} \, \mbox{\boldmath $x$} \ ,
\end{equation}
where detector coefficients {\boldmath $f$} and pixel terms
{\boldmath $x$} partition into rank 1 and rank 2 tensors as
\begin{equation}
\mbox{\boldmath $f$} \equiv \left[
\begin{array}{c}
\mbox{\boldmath $f$}^{(1)} \\
\mbox{\boldmath $f$}^{(2)}
\end{array}
\right] \ , \ \mbox{\boldmath $x$} \equiv \left[
\begin{array}{c}
\mbox{\boldmath $x$}^{(1)} \\
\mbox{\boldmath $x$}^{(2)}
\end{array}
\right] \ .
\end{equation}
Pixel terms have mean
\begin{equation}
\begin{array}{l}
\mbox{\boldmath $g$} = \left[
\begin{array}{c}
\mbox{\boldmath $g$}^{(1)} \\* \mbox{\boldmath $g$}^{(2)}
\end{array}
\right] \equiv \langle \mbox{\boldmath $x$} \rangle \; : \\*[3ex]
\mbox{\boldmath $g$}^{(i)} \equiv \langle \mbox{\boldmath
$x$}^{(i)} \rangle \ ,
\end{array}
\end{equation}
and covariance
\begin{equation}
\begin{array}{l}
C = \left[
\begin{array}{cc}
C^{(2)} & C^{(1+2)} \\* C^{(2+1)} & C^{(4)}
\end{array}
\right] \equiv \left \langle (\mbox{\boldmath $x$} - \langle
\mbox{\boldmath $x$} \rangle) (\mbox{\boldmath $x$} - \langle
\mbox{\boldmath $x$} \rangle)^{\mathrm{T}} \right \rangle \; :
\\*[3ex]
C^{(i+j)} \equiv \left \langle (\mbox{\boldmath
$x$}^{(i)} - \langle \mbox{\boldmath $x$}^{(i)} \rangle)
(\mbox{\boldmath $x$}^{(j)} - \langle \mbox{\boldmath $x$}^{(j)}
\rangle)^{\mathrm{T}} \right \rangle \ ,
\end{array}
\end{equation}
where (as usual) superscripts of the form $(i)$ indicate that the
quantity is inherently a rank~$i$ tensor.

\begin{table*}[t!]
\begin{center}
{\footnotesize
\begin{tabular}{|r|r|rrr|rrr|rrr|}
\hline

 &
 &
\multicolumn{3}{l|}{ } & \multicolumn{3}{l|}{Degrees of
freedom---} & \multicolumn{3}{l|}{Degrees of freedom---}
\\
 &
 &
\multicolumn{3}{l|}{ } & \multicolumn{3}{l|}{permutation symmetry}
& \multicolumn{3}{l|}{permutation and}
\\
 &
 &
\multicolumn{3}{l|}{Detector coefficients:} &
\multicolumn{3}{l|}{only:} & \multicolumn{3}{l|}{geometric
symmetry:}
\\
\multicolumn{1}{|c|}{$n$} & \multicolumn{1}{c|}{$N$} &
\multicolumn{1}{c}{linear} & \multicolumn{1}{c}{quadratic} &
\multicolumn{1}{c|}{total} & \multicolumn{1}{c}{linear} &
\multicolumn{1}{c}{quadratic} & \multicolumn{1}{c|}{total} &
\multicolumn{1}{c}{linear} & \multicolumn{1}{c}{quadratic} &
\multicolumn{1}{c|}{total}
\\

\hline \hline

      1 &      1 &      1 &      1 &      2 &      1 &      1 &      2 &
      1 &      1 &      2 \\

      2 &      4 &      4 &     16 &     20 &      4 &     10 &     14 &
      1 &      3 &      4 \\

      3 &      9 &      9 &     81 &     90 &      9 &     45 &     54 &
      3 &     11 &     14 \\

      4 &     16 &     16 &    256 &    272 &     16 &    136 &    152 &
      3 &     24 &     27 \\

      5 &     25 &     25 &    625 &    650 &     25 &    325 &    350 &
      6 &     55 &     61 \\

      6 &     36 &     36 &   1296 &   1332 &     36 &    666 &    702 &
      6 &     99 &    105 \\

      7 &     49 &     49 &   2401 &   2450 &     49 &   1225 &   1274 &
     10 &    181 &    191 \\

      8 &     64 &     64 &   4096 &   4160 &     64 &   2080 &   2144 &
     10 &    288 &    298 \\

      9 &     81 &     81 &   6561 &   6642 &     81 &   3321 &   3402 &
     15 &    461 &    476 \\

     10 &    100 &    100 &  10000 &  10100 &    100 &   5050 &   5150 &
     15 &    675 &    690 \\

     12 &    144 &    144 &  20736 &  20880 &    144 &  10440 &  10584 &
     21 &   1368 &   1389 \\

     14 &    196 &    196 &  38416 &  38612 &    196 &  19306 &  19502 &
     28 &   2499 &   2527 \\

     16 &    256 &    256 &  65536 &  65792 &    256 &  32896 &  33152 &
     36 &   4224 &   4260 \\

     18 &    324 &    324 & 104976 & 105300 &    324 &  52650 &  52974 &
     45 &   6723 &   6768 \\

     20 &    400 &    400 & 160000 & 160400 &    400 &  80200 &  80600 &
     55 &  10200 &  10255 \\

     25 &    625 &    625 & 390625 & 391250 &    625 & 195625 & 196250 &
     91 &  24805 &  24896 \\

\hline
\end{tabular}}
\end{center}
\caption{(Refer to Section~\ref{sec: D matrix}) Square lattice.
Reduction in degrees of freedom due to permutation symmetry and
geometric symmetry. Support linear dimension is~$n$, and number of
pixels is~$N$. The numbers of detector coefficients and degrees of
freedom are resolved into linear and quadratic types.}
\label{tab:sq_dof}
\end{table*}

Ensemble statistics are invariant with respect to all symmetry
transformations if and only if
\begin{eqnarray}
\label{statistics invariance}
 & \mbox{\boldmath $g$}^{(i)} =
T^{(2i)}_{m} \, \mbox{\boldmath $g$}^{(i)} \ \ \mbox{and} \ \
C^{(i+j)} = T^{(2i)}_{m} \, C^{(i+j)} \, T^{(2j)\,\mathrm{T}}_{m}
\, , & \nonumber \\
 & \hspace{-10em} m=1,\dots,s \ . &
\end{eqnarray}
Using the fact that any group element multiplying separately all
group elements (including itself) yields all the group elements in
a different order, one may verify that
\begin{equation}
\label{g and C tilde def}
\begin{array}{l}
\widetilde{\mbox{\boldmath $g$}}^{(i)} \equiv {\displaystyle
\frac{1}{s} \sum_{m=1}^{s}} T^{(2i)}_{m} \, \mbox{\boldmath
$g$}^{(i)} \ , \\*[4ex] \widetilde{C}^{(i+j)} \equiv
{\displaystyle \frac{1}{s} \sum_{m=1}^{s}} T^{(2i)}_{m} \,
C^{(i+j)} \, T^{(2j)\,\mathrm{T}}_{m} \, ,
\end{array}
\end{equation}
satisfy invariance properties (\ref{statistics invariance}) for
arbitrary {\boldmath $g$}$^{(i)}$ and $C^{(i+j)}$. Furthermore,
$\widetilde{\mbox{\boldmath $g$}}^{(i)}$ and
$\widetilde{C}^{(i+j)}$ have the correct form to be ensemble
statistics. Although $(\sum_{m} T^{(2i)}_{m}) C^{(i+j)} (\sum_{n}
T^{(2j)}_{n})$ has the correct invariance property, it is a
cross-covariance between all pairs of symmetry transformed images,
whereas a suitable $\widetilde{C}^{(i+j)}$ must be an
autocovariance of all symmetry transformed images. The $1/s$
normalising factors in (\ref{g and C tilde def}) ensure that if
{\boldmath $g$}$^{(i)}$ and $C^{(i+j)}$ already satisfy statistics
invariance conditions (\ref{statistics invariance}), then
resymmetrisation by (\ref{g and C tilde def}) will leave
{\boldmath $g$}$^{(i)}$ and $C^{(i+j)}$ unchanged. The QFD
analysis proceeds with symmetrised ensemble statistics
$\widetilde{\mbox{\boldmath $g$}}^{(i)}$ and
$\widetilde{C}^{(i+j)}$ in place of their unsymmetrised
counterparts {\boldmath $g$}$^{(i)}$ and $C^{(i+j)}$.

The QFD is the quadratic filter whose output maximises the ratio
of the squared difference of means of targets and clutter
(dividend), to the sum of variances of targets and clutter
(divisor). This QFD objective function is
\begin{equation}
\label{objective function 1} \sigma(\mbox{\boldmath $f$}) =
\frac{\left( \mbox{\boldmath $f$}^{\mathrm{T}}
\widetilde{\mbox{\boldmath $g$}} \right)^{2}}{ \mbox{\boldmath
$f$}^{\mathrm{T}} \widetilde{C} \mbox{\boldmath $f$}} \ ,
\end{equation}
which is maximised by the solution to
\begin{equation}
\label{Cf=g 1} \widetilde{C} \mbox{\boldmath $f$} =
\widetilde{\mbox{\boldmath $g$}} \ .
\end{equation}
In (\ref{objective function 1}) and (\ref{Cf=g 1})
$\widetilde{\mbox{\boldmath $g$}}$ is the difference of means, and
$\widetilde{C}$ is the sum of covariances, for clutter and target
ensembles:
\begin{equation}
\begin{array}{rcl}
\widetilde{\mbox{\boldmath $g$}} & \equiv &
\widetilde{\mbox{\boldmath $g$}}_{\mathrm{clutter}} -
\widetilde{\mbox{\boldmath $g$}}_{\mathrm{target}} \ , \\*[2ex]
\widetilde{C} & \equiv & \widetilde{C}_{\mathrm{clutter}} +
\widetilde{C}_{\mathrm{target}} \ . \\
\end{array}
\end{equation}
Ensemble statistics invariance properties (\ref{statistics
invariance}) applied to (\ref{Cf=g 1}) derive detector coefficient
invariance properties
\begin{equation}
\label{detector invariance} \mbox{\boldmath $f$}^{(i)} =
T^{(2i)}_{m} \, \mbox{\boldmath $f$}^{(i)} \, , \ m=1,\dots,s \ ,
\end{equation}
under the assumption that $\widetilde{C}$ is nonsingular and so
(\ref{Cf=g 1}) has a unique solution\footnote{$\widetilde{C}$
actually has essential singularities due to permutation symmetry,
but their removal will be addressed soon.}. For the same reasons
that apply to (\ref{g and C tilde def}),
\begin{equation}
\label{symmetrised detector} \widetilde{\mbox{\boldmath
$f$}}^{(i)} = \frac{1}{s} \sum_{m=1}^{s} T^{(2i)}_{m} \,
\mbox{\boldmath $f$}^{(i)} \ ,
\end{equation}
satisfies invariance property (\ref{detector invariance}) for
arbitrary {\boldmath $f$}$^{(i)}$. Detector coefficients
invariance (\ref{detector invariance}) applied to (\ref{detector
operation}) yields the detector response invariance property
\begin{equation}
\label{response invariance} y(T^{(2i)}_{m} \, \mbox{\boldmath
$x$}^{(i)}) = y(\mbox{\boldmath $x$}^{(i)}) \, , \ m=1,\dots,s \ ,
\end{equation}
which implies that the detector response is the same for all
images that are point symmetry transformations of each other.
Unlike this work, most approaches to filter
symmetrisation~\cite{Giles87}--\cite{Sams00} postulate detector
response symmetry, from which detector coefficient symmetry
follows.

If the argument of objective function $\sigma$ (\ref{objective
function 1}) is restricted to symmetrised detector coefficients
$\widetilde{\mbox{\boldmath $f$}}$ (\ref{symmetrised detector}),
then explicit symmetrisation of ensemble statistics becomes
redundant in the computation of $\sigma$, that is,
\begin{equation}
\label{objective function 2} \sigma(\widetilde{\mbox{\boldmath
$f$}}) = \frac{\left( \widetilde{\mbox{\boldmath
$f$}}^{\mathrm{T}} \widetilde{\mbox{\boldmath $g$}} \right)^{2}}{
\widetilde{\mbox{\boldmath $f$}}^{\mathrm{T}} \widetilde{C}
\widetilde{\mbox{\boldmath $f$}}} = \frac{\left(
\widetilde{\mbox{\boldmath $f$}}^{\mathrm{T}} \mbox{\boldmath $g$}
\right)^{2}}{ \widetilde{\mbox{\boldmath $f$}}^{\mathrm{T}} C
\widetilde{\mbox{\boldmath $f$}}} \ .
\end{equation}

\section{Degeneracy matrix}
\label{sec: D matrix}

Solving (\ref{Cf=g 1}) directly yields detector coefficients that
are properly symmetrised to within numerical accuracy. Then if
necessary, explicit symmetrisation of the detector coefficients by
(\ref{symmetrised detector}) will give perfectly symmetrised
detector coefficients. However, there is a more elegant and
computationally efficient method of computing perfectly
symmetrised detector coefficients. This superior method introduces
a degeneracy\footnote{A set of quantities is said to be
\emph{degenerate} if all of the quantities are identically equal,
and not simply equal by accident. The concepts of degeneracy and
symmetry are synonymous.} matrix to account for permutation and
geometric symmetry degeneracies.

As a consequence of the group properties satisfied by matrices
$T^{(2i)}_{m}$, matrix
\begin{equation}
D^{(2i)\prime\prime\prime} \equiv \sum_{m=1}^{s} T^{(2i)}_{m} \ ,
\end{equation}
satisfies the symmetry invariance properties
\begin{equation}
D^{(2i)\prime\prime\prime} = T^{(2i)}_{m} \,
D^{(2i)\prime\prime\prime} \, , \ m=1,\dots,s \ .
\end{equation}
Matrix $D^{(2i)\prime\prime\prime}$ is symmetric,
\begin{equation}
D^{(2i)\prime\prime\prime\,\mathrm{T}} =
D^{(2i)\prime\prime\prime} \ ,
\end{equation}
due to the orthogonality property (\ref{matrix orthogonality}) of
matrices $T^{(2i)}_{m}$. $D^{(2i)\prime\prime\prime}$ has nonzero
entries everywhere along the main diagonal. If column $i_{1}$ has
nonzero entries in rows $i_{1},i_{2},i_{3},\ldots$ (maximum of $s$
distinct rows), then columns $i_{1},i_{2},i_{3},\ldots$ are
identical. The nonzero entries in a given column all have the same
value and add to $s$\footnote{\label{fn: D''' columns}A square
lattice has $s\!=\!8$, and columns of $D^{(2i)\prime\prime\prime}$
may have 8 1s, or 4 2s, or 2 4s, or 1 8 entry, although only
$D^{(4)\prime\prime\prime}$ actually has columns with 2 4
entries.}.

Matrix $D^{(2i)\prime\prime}$ has a one entry wherever
$D^{(2i)\prime\prime\prime}$ has any nonzero entry:
\begin{equation}
d^{(2i)\prime\prime}(k^{(i)};l^{(i)}) \equiv \left\{
\begin{array}{l}
1 \ \mbox{if} \ d^{(2i)\prime\prime\prime}(k^{(i)};l^{(i)})\neq0 \\
0 \ \mbox{otherwise}
\end{array}
\right. \ ,
\end{equation}
where the notation $k^{(i)}$ is an abbreviation for a
$2i$-component index. Matrix $D^{(2i)\prime\prime}$ is symmetric.
Because each matrix $T^{(2i)}_{m}$ contains only a single one
entry per row or column, and zeros elsewhere,
$D^{(2i)\prime\prime}$ satisfies the symmetry invariance
properties
\begin{equation}
D^{(2i)\prime\prime} = T^{(2i)}_{m} \, D^{(2i)\prime\prime} \, , \
m=1,\dots,s \ .
\end{equation}

Additional to the geometric symmetry being postulated in this
analysis, it is convenient to impose the permutation symmetry
\begin{equation}
\label{permutation symmetry} f^{(2)}(\{ij,kl\}) =
f^{(2)}(\{kl,ij\})
\end{equation}
on the rank 2 tensor detector coefficients. Permutation symmetry
(\ref{permutation symmetry}) of {\boldmath $f$}$^{(2)}$ is chosen
to correspond to the permutation symmetry of {\boldmath
$x$}$^{(2)}$ (\ref{x_2 components}). Matrix $D^{(4)\prime}$ has a
one entry if $D^{(4)\prime\prime}$ has a one entry at the same row
and column, or the same row but permuted column 4-index:
\begin{eqnarray}
\lefteqn{d^{(4)\prime}(\{ij,kl\};\{i'j',k'l'\}) \equiv} \nonumber
\\[1ex]
 & & \left\{
\begin{array}{l}
1 \ \mbox{if} \ d^{(4)\prime\prime}(\{ij,kl\};\{i'j',k'l'\})=1
\ \mbox{or} \\
\phantom{1 \ \mbox{if} \ }
d^{(4)\prime\prime}(\{ij,kl\};\{k'l',i'j'\})=1 \\[1ex]
0 \ \mbox{otherwise}
\end{array}
\right. \ .
\end{eqnarray}
$D^{(4)\prime}$ satisfies the geometric symmetry invariance
properties
\begin{equation}
D^{(4)\prime} = T^{(4)}_{m} \, D^{(4)\prime} \, , \ m=1,\dots,s \
,
\end{equation}
and the permutation symmetry invariance properties
\begin{equation}
\label{D4 perm sym}
\begin{array}{l}
d^{(4)\prime}(\{ij,kl\};\{i'j',k'l'\}) =
d^{(4)\prime}(\{ij,kl\};\{k'l',i'j'\}) = \\[1ex]
d^{(4)\prime}(\{kl,ij\};\{i'j',k'l'\}) =
d^{(4)\prime}(\{kl,ij\};\{k'l',i'j'\}) \ .
\end{array}
\end{equation}
Additional to (\ref{D4 perm sym}) is the property that
$D^{(4)\prime}$ is a symmetric matrix, which may be regarded as
another, unavoidable permutation symmetry. The redefinition
\begin{equation}
D^{(2)\prime} \equiv D^{(2)\prime\prime} \ ,
\end{equation}
is adopted for notational consistency. Covariance matrix $C$ and
its derivatives have multiple essential singularities if
permutation symmetry is not explicitly taken into account, as is
done here by replacing $D^{(2i)\prime\prime}$ by $D^{(2i)\prime}$.

\begin{figure}[t!]
\begin{center}
\includegraphics{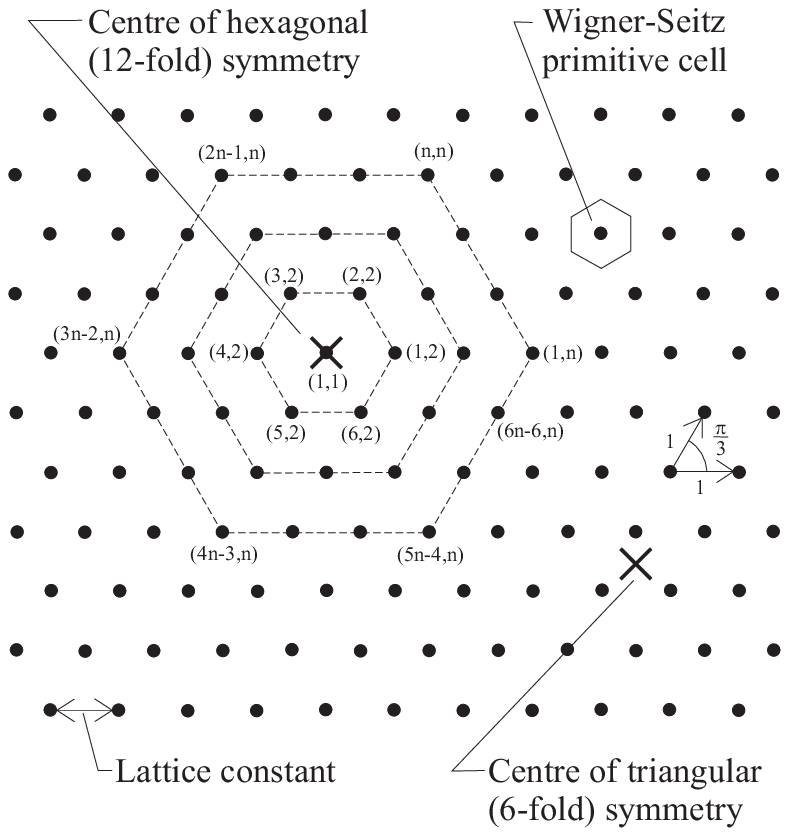}
\end{center}
\caption{(Refer to Section~\ref{sec: hexagonal lattice}) A section
of hexagonal lattice. Nonequivalent points marked $\times$
correspond to different symmetries---lattice points have hexagonal
symmetry, and interstitial sites have triangular symmetry. Shown
is a detector support of linear dimension~$n$, and the pixel
indexing scheme.} \label{fig: hex_net}
\end{figure}

Define the reduced width matrix $D^{(2i)}$ as having columns that
are all of the distinct columns of matrix $D^{(2i)\prime}$,
without repetition. Repeated columns in $D^{(2i)\prime}$ are
identified by the property that if column $i_{1}$ has one entries
in rows $i_{1},i_{2},i_{3},\ldots$\footnote{Maximum of $s$
distinct rows for $D^{(2)\prime}$; maximum of $2s$ distinct rows
for $D^{(4)\prime}$.}, then columns $i_{1},i_{2},i_{3},\ldots$ are
identical, and there are no other columns that are identical to
column $i_{1}$. This property derives from the analogous property
of $D^{(2i)\prime\prime\prime}$ that was noted earlier in this
section. Matrices $D^{(2)}$ and $D^{(4)}$ respectively are the
rank 2 and rank 4 tensor\footnote{Since the column index of
$D^{(2i)}$ can not be expressed in terms of 2-indices, $D^{(2i)}$
are strictly not tensors in the context of this analysis. But
matrices $D^{(2i)}$ are derived from rank $2i$ tensors, hence the
terminology.} parts of degeneracy matrix $D$:
\begin{equation}
D \equiv \left[
\begin{array}{cc}
D^{(2)} & 0 \\
0 & D^{(4)}
\end{array}
\right] \ .
\end{equation}
$D^{(2i)}$ satisfies the geometric symmetry invariance properties
\begin{equation}
\label{D geo sym} D^{(2i)} = T^{(2i)}_{m} \, D^{(2i)} \, , \
m=1,\dots,s \ ,
\end{equation}
and the permutation symmetry invariance properties
\begin{equation}
d^{(4)}(\{ij,kl\};m) = d^{(4)}(\{kl,ij\};m) \ .
\end{equation}

\begin{table*}[t!]
\begin{center}
{\footnotesize
\begin{tabular}{|l|l|l|l|}
\hline \rule{0em}{2.5ex}

Symmetry: & Description of symmetry & Factorisation &
Rank 2 tensor representation:   \\
$T_{m}$   & operation               &               &
$t^{(2)}_{m}(\{ij\};\{i'j'\})$ \\[0.5ex]

\hline \hline \rule{0em}{2.5ex}

$T_{1}$ & Identity or unit operation.  & $T_{7}^{\,2}$      &
$\delta(i,i').$ \\
        &                              &                  &
$\delta(j,j')$  \\[0.5ex]

\hline \rule{0em}{2.5ex}

$T_{2}$ & Anticlockwise rotation by    & $T_{2}$          &
$\delta(i,(i'+j'-2) \bmod 6(j'-1)+1).$ \\
        & 1/6 turn.                    &                  &
$\delta(j,j')$ \\[0.5ex]

\hline \rule{0em}{2.5ex}

$T_{3}$ & Anticlockwise rotation by    & $T_{2}^{\,2}$      &
$\delta(i,(i'+2j'-3) \bmod 6(j'-1)+1).$ \\
        & 1/3 turn.                    &                  &
$\delta(j,j')$ \\[0.5ex]

\hline \rule{0em}{2.5ex}

$T_{4}$ & Rotation by 1/2 turn, being  & $T_{2}^{\,3}$      &
$\delta(i,(i'+3j'-4) \bmod 6(j'-1)+1).$ \\
        & inversion through centre.    &                  &
$\delta(j,j')$ \\[0.5ex]

\hline \rule{0em}{2.5ex}

$T_{5}$ & Clockwise rotation by 1/3    & $T_{2}^{\,4}$      &
$\delta(i,(i'+4j'-5) \bmod 6(j'-1)+1).$ \\
        & turn.                        &                  &
$\delta(j,j')$  \\[0.5ex]

\hline \rule{0em}{2.5ex}

$T_{6}$ & Clockwise rotation by 1/6    & $T_{2}^{\,5}$      &
$\delta(i,(i'+5j'-6) \bmod 6(j'-1)+1).$ \\
        & turn.                        &                  &
$\delta(j,j')$  \\[0.5ex]

\hline \rule{0em}{2.5ex}

$T_{7}$ & Reflection in the $x$-axis.  & $T_{7}$          &
$\delta(i,(6j'-i'-5) \bmod 6(j'-1)+1).$ \\
        &                              &                  &
$\delta(j,j')$ \\[0.5ex]

\hline \rule{0em}{2.5ex}

$T_{8}$ & Reflection in line 1/12 turn & $T_{2}\,T_{7}$     &
$\delta(i,(7j'-i'-6) \bmod 6(j'-1)+1).$ \\
        & anticlockwise from $x$-axis. &                  &
$\delta(j,j')$ \\[0.5ex]

\hline \rule{0em}{2.5ex}

$T_{9}$ & Reflection in line 1/6 turn  & $T_{2}^{\,2}\,T_{7}$
&
$\delta(i,(8j'-i'-7) \bmod 6(j'-1)+1).$ \\
        & anticlockwise from $x$-axis. &                  &
$\delta(j,j')$ \\[0.5ex]

\hline \rule{0em}{2.5ex}

$T_{10}$ & Reflection in the $y$-axis.  & $T_{2}^{\,3}\,T_{7}$ &
$\delta(i,(9j'-i'-8) \bmod 6(j'-1)+1).$ \\
        &                              &                  &
$\delta(j,j')$ \\[0.5ex]

\hline \rule{0em}{2.5ex}

$T_{11}$ & Reflection in line 1/6 turn  & $T_{2}^{\,4}\,T_{7}$ &
$\delta(i,(10j'-i'-9) \bmod 6(j'-1)+1).$ \\
        & clockwise from $x$-axis.     &                  &
$\delta(j,j')$ \\[0.5ex]

\hline \rule{0em}{2.5ex}

$T_{12}$ & Reflection in line 1/12 turn & $T_{2}^{\,5}\,T_{7}$ &
$\delta(i,(11j'-i'-10) \bmod 6(j'-1)+1).$ \\
        & clockwise from $x$-axis.     &                  &
$\delta(j,j')$ \\[0.5ex]

\hline
\end{tabular}}
\end{center}
\caption{(Refer to Section~\ref{sec: hexagonal lattice}) Hexagonal
lattice point group symmetry operations and their rank 2 tensor
representation; the rank 4 tensor representation is derived from
the rank 2 tensor representation according to (\ref{t(4) from
t(2)}). Orientation of the lattice is such that a basis vector
aligns with the $x$ axis, as in Figure~\ref{fig: hex_net}. Note
that the modulo operation is always with respect to $6(j'-1)$.}
\label{tab:hex_point_group}
\end{table*}

Let $\widehat{\mbox{\boldmath $f$}}$ be an unconstrained
nondegenerate detector coefficient vector. Then
\begin{equation}
\label{f-hat} \widetilde{\mbox{\boldmath $f$}} = D
\widehat{\mbox{\boldmath $f$}}
\end{equation}
is a degenerate coefficient vector that has both the required
geometric symmetry (\ref{detector invariance}) and permutation
symmetry (\ref{permutation symmetry}). Each row of matrix $D$ has
a single one entry and zeros elsewhere. Multiple one entries in a
column of $D$ associate degenerate coefficients in the symmetrised
detector $\widetilde{\mbox{\boldmath $f$}}$. The number of columns
in $D$, and the number of components in $\widehat{\mbox{\boldmath
$f$}}$, is the number of degrees of freedom\footnote{\emph{Degrees
of freedom} of the detector are the independent coefficients
remaining after accounting for all symmetries and before
maximising the objective function.} of the detector. Substituting
(\ref{f-hat}) into (\ref{objective function 2}) re-expresses the
objective function as
\begin{equation}
\label{objective function 3} \sigma(\widehat{\mbox{\boldmath
$f$}}) = \frac{\left(\widehat{\mbox{\boldmath $f$}}^{\mathrm{T}}
\widehat{\mbox{\boldmath $g$}} \right)^{2}}{
\widehat{\mbox{\boldmath $f$}}^{\mathrm{T}} \widehat{C}
\widehat{\mbox{\boldmath $f$}}} \ ,
\end{equation}
where the nondegenerate mean vector $\widehat{\mbox{\boldmath
$g$}}$ is
\begin{equation}
\label{g-hat def} \widehat{\mbox{\boldmath $g$}} \equiv D^{\mathrm
T} \widetilde{\mbox{\boldmath $g$}} = D^{\mathrm T}
\mbox{\boldmath $g$} \ ,
\end{equation}
and the nondegenerate covariance matrix $\widehat{C}$ is
\begin{equation}
\label{C-hat def} \widehat{C} \equiv D^{\mathrm T} \widetilde{C} D
= D^{\mathrm T} C D \ .
\end{equation}
The second equalities in (\ref{g-hat def}) and (\ref{C-hat def}),
which also may be derived from (\ref{g and C tilde def}) and
(\ref{D geo sym}), affirm the redundancy of statistics
symmetrisation already noted in association with (\ref{objective
function 2}). Equation (\ref{C-hat def}) shows that the degeneracy
matrix formalism maximally compresses the covariance matrix
without loss of information; an alternative approach due to
Lenz~\cite{Lenz93,Lenz95} partially compresses the covariance
matrix into an equal size block diagonal matrix, whose top-left
block has precisely the same size as $\widehat{C}$ of~(\ref{C-hat
def}).

Maximisation of objective function (\ref{objective function 3}) is
equivalent to solving (cf. (\ref{Cf=g 1}))
\begin{equation}
\label{Cf=g: carets} \widehat{C} \widehat{\mbox{\boldmath $f$}} =
\widehat{\mbox{\boldmath $g$}} \ ,
\end{equation}
whose solution is
\begin{equation}
\label{generalised inverse} \widehat{\mbox{\boldmath $f$}} =
\widehat{C}^{\dag} \widehat{\mbox{\boldmath $g$}} \ ,
\end{equation}
where $\widehat{C}^{\dag}$ is the Moore-Penrose generalised
inverse (or pseudoinverse) \cite{Penrose55} of
$\widehat{\mbox{$C$}}$. A property of solutions of the form
(\ref{generalised inverse}) obtained by generalised inversion is
that they are the minimum norm vector that minimises the Euclidean
norm of the residual $( \widehat{C} \widehat{\mbox{\boldmath $f$}}
- \widehat{\mbox{\boldmath $g$}} )$ \cite{Penrose56}. The residual
vanishes if $\widehat{\mbox{$C$}}$ is nonsingular, in which case
the generalised inverse is equivalent to the conventional inverse.
QFD generalisation\footnote{\emph{Generalisation} of a detector
refers to the consistency of detector performance when assessed
separately against training and test ensembles.} may be improved
by regularising $\widehat{\mbox{$C$}}$~\cite{Friedman89, Hastie95}
before solving~(\ref{Cf=g: carets}).

A necessary and sufficient condition for $\widehat{C}$ to be
nonsingular is that the clutter and target training ensembles
between them must contain no fewer linearly independent images
than the number of symmetrised detector degrees of freedom. In
contrast, symmetrised covariance matrix $\widetilde{C}$ (and
unsymmetrised covariance matrix $C$) contains many essential
singularities, no matter how large the training ensembles. Zero
eigenvalues of $\widetilde{C}$ are at least equal in number to the
number of permutation symmetry degeneracies in the detector
coefficients; beyond that there may be extra zero eigenvalues due
to small training ensemble sizes. The present degeneracy matrix
removes the essential singularities in $\widetilde{C}$ or $C$ by
operation~(\ref{C-hat def}).

Table~\ref{tab:sq_dof} shows the reduction in degrees of freedom
afforded by permutation and geometric symmetries. Permutation
symmetry among the $N\!+\!N^{2}$ coefficients leaves
$N\!+\!N(N\!+\!1)/2$ degrees of freedom. This is a reduction in
degrees of freedom by a factor of up to 2. Maximum reduction is
achieved for large detector supports, where the proportion of
quadratic terms that are pixel self-products becomes small.
Geometric symmetry further reduces the degrees of freedom by a
factor of up to $s$ ($s\!=\!8$ for a square lattice, to which
Table~\ref{tab:sq_dof} applies). Maximum reduction occurs for
large detector supports, where proportionately few pixels and
pixel pairs map onto either themselves under symmetry
transformations, or common pixels or pixel pairs under different
symmetry transformations. The reduction in degrees of freedom
attributable to geometric symmetry should constitute a significant
improvement in detector generalisation, when the unsymmetrised
detector\footnote{That is, a detector that accounts for
permutation symmetry, but not geometric symmetry.} generalises
poorly.

\section{Hexagonal lattice symmetry}
\label{sec: hexagonal lattice}

\begin{table*}[t!]
\begin{center}
{\footnotesize
\begin{tabular}{|r|r|rrr|rrr|rrr|}
\hline

 &
 &
\multicolumn{3}{l|}{ } & \multicolumn{3}{l|}{Degrees of
freedom---} & \multicolumn{3}{l|}{Degrees of freedom---}
\\
 &
 &
\multicolumn{3}{l|}{ } & \multicolumn{3}{l|}{permutation symmetry}
& \multicolumn{3}{l|}{permutation and}
\\
 &
 &
\multicolumn{3}{l|}{Detector coefficients:} &
\multicolumn{3}{l|}{only:} & \multicolumn{3}{l|}{geometric
symmetry:}
\\
\multicolumn{1}{|c|}{$n$} & \multicolumn{1}{c|}{$N$} &
\multicolumn{1}{c}{linear} & \multicolumn{1}{c}{quadratic} &
\multicolumn{1}{c|}{total} & \multicolumn{1}{c}{linear} &
\multicolumn{1}{c}{quadratic} & \multicolumn{1}{c|}{total} &
\multicolumn{1}{c}{linear} & \multicolumn{1}{c}{quadratic} &
\multicolumn{1}{c|}{total}
\\

\hline \hline

      1 &      1 &      1 &      1 &      2 &      1 &      1 &      2 &
      1 &      1 &      2 \\

      2 &      7 &      7 &     49 &     56 &      7 &     28 &     35 &
      2 &      6 &      8 \\

      3 &     19 &     19 &    361 &    380 &     19 &    190 &    209 &
      4 &     26 &     30 \\

      4 &     37 &     37 &   1369 &   1406 &     37 &    703 &    740 &
      6 &     77 &     83 \\

      5 &     61 &     61 &   3721 &   3782 &     61 &   1891 &   1952 &
      9 &    189 &    198 \\

      6 &     91 &     91 &   8281 &   8372 &     91 &   4186 &   4277 &
     12 &    394 &    406 \\

      7 &    127 &    127 &  16129 &  16256 &    127 &   8128 &   8255 &
     16 &    742 &    758 \\

      8 &    169 &    169 &  28561 &  28730 &    169 &  14365 &  14534 &
     20 &   1281 &   1301 \\

      9 &    217 &    217 &  47089 &  47306 &    217 &  23653 &  23870 &
     25 &   2081 &   2106 \\

     10 &    271 &    271 &  73441 &  73712 &    271 &  36856 &  37127 &
     30 &   3206 &   3236 \\

     11 &    331 &    331 & 109561 & 109892 &    331 &  54946 &  55277 &
     36 &   4746 &   4782 \\

     12 &    397 &    397 & 157609 & 158006 &    397 &  79003 &  79400 &
     42 &   6781 &   6823 \\

     13 &    469 &    469 & 219961 & 220430 &    469 & 110215 & 110684 &
     49 &   9421 &   9470 \\

     14 &    547 &    547 & 299209 & 299756 &    547 & 149878 & 150425 &
     56 &  12762 &  12818 \\

     15 &    631 &    631 & 398161 & 398792 &    631 & 199396 & 200027 &
     64 &  16934 &  16998 \\

\hline
\end{tabular}}
\end{center}
\caption{(Refer to Section~\ref{sec: hexagonal lattice}) Hexagonal
lattice. Reduction in degrees of freedom due to permutation
symmetry and geometric symmetry. Support linear dimension is $n$,
and number of pixels is $N$. The numbers of detector coefficients
and degrees of freedom are resolved into linear and quadratic
types.} \label{tab:hex_dof}
\end{table*}

The most symmetric possible plane lattice has a 6-fold rotation
axis through a lattice point, around which are 6 mirror lines in
the lattice plane uniformly spaced by angle $\pi/6$; this lattice
is the hexagonal lattice\footnote{The hexagonal lattice is often
called the triangular lattice, since interstices have three
lattice points at the vertices of equilateral triangles as nearest
neighbours, much as square lattice interstices have four lattice
points at the vertices of squares as nearest neighbours. The
alternative convention used here is that lattices are named
according to their point symmetry and their Wigner-Seitz primitive
cell.}. As shown in Figure~\ref{fig: hex_net}, the hexagonal
lattice is generated by two equal length basis vectors subtending
angle $\pi/3$. With centre of symmetry at a lattice point,
hexagonal lattice point symmetry is of order
\begin{equation}
s = 12 \hspace{2.0em} \mbox{(hexagonal lattice)} \ .
\end{equation}
Unlike the square lattice, interstitual sites in the hexagonal
lattice have reduced symmetry compared with lattice point centres.
The 6-fold point symmetry of hexagonal lattice
interstices\footnote{Unlike square lattices, hexagonal lattice
interstices are of two nonequivalent types, depending on whether
they are centred on upright or inverted triangles of nearest
neighbour lattice points.} is even lower than the 8-fold point
symmetry of square lattices. Clearly, one would not sample on an
hexagonal lattice, only to use detector supports centred on
interstices. Having established that the detector support will be
centred only on lattice points, Figure~\ref{fig: hex_net} shows a
regular hexagonal support of linear dimension $n$, and a suitable
pixel indexing scheme. The support linear dimension $n$ is the
number of concentric hexagons that account for all pixels, where
the central lattice point counts as a hexagon of vanishing size;
equivalently, the linear dimension is the number of lattice points
between adjacent vertices of the support perimeter. A spiral
indexing scheme is adopted for the hexagonal lattice, where
$(i,j)$ represents the $i$th lattice point in the $j$th concentric
hexagon. An analogous spiral indexing scheme could have been used
for the square lattice, the result being a closer correspondence
between the point symmetry tensor representations for square and
hexagonal lattices than is evident in comparing Tables
\ref{tab:sq_point_group} and \ref{tab:hex_point_group}. The
row-column indexing scheme actually used for the square lattice in
Table \ref{tab:sq_point_group} is conventional, and allows simpler
expression of tensor representations, as is evident in comparing
Tables \ref{tab:sq_point_group} and \ref{tab:hex_point_group}.

\begin{figure}[t!]
\begin{center}
\includegraphics{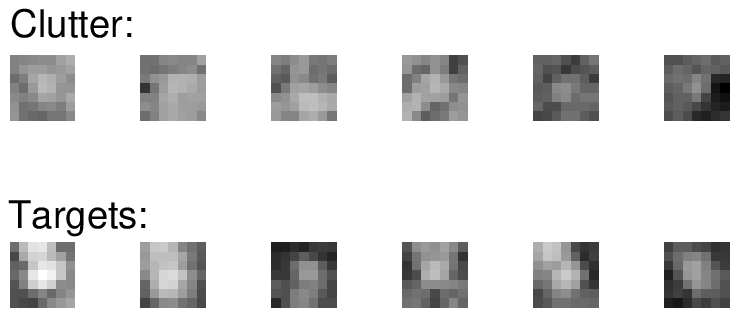}
\end{center}
\caption{(Refer to Section~\ref{sec: experiment}) Sample clutter
and target $9\!\times\!9$ image chips as used in the experiment.}
\label{fig:image chips}
\end{figure}

\begin{figure*}[t!]
\begin{center}
\includegraphics{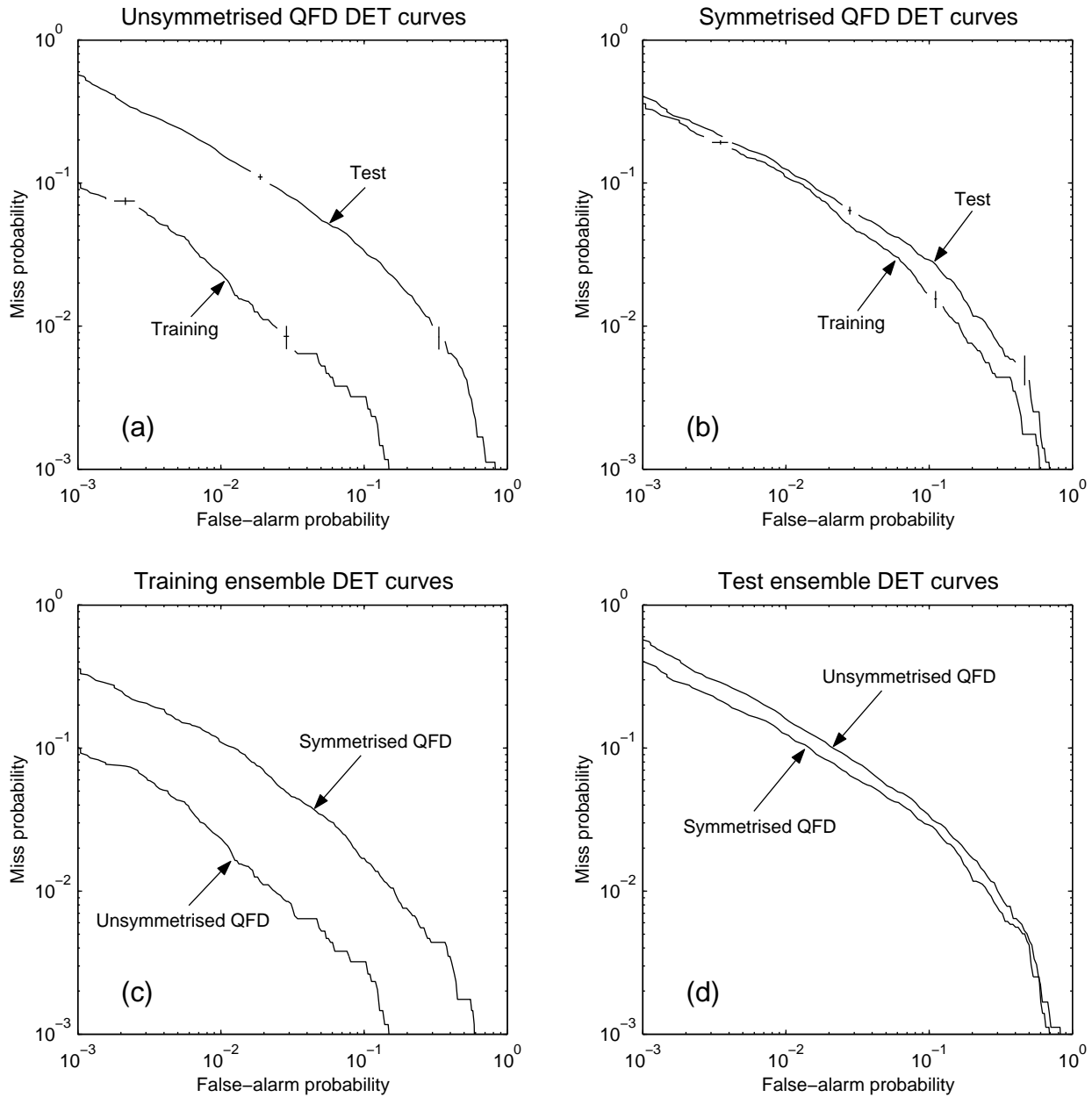}
\end{center}
\caption{(Refer to Section~\ref{sec: experiment}) Training and
test ensemble DET curves of the unsymmetrised and symmetrised
QFDs, plotted in various combinations to aid comparison. Dual axis
error bars are shown in graphs (a) and~(b).} \label{fig:exp_res}
\end{figure*}

The square lattice analysis of Section~\ref{sec: square lattice}
also holds for the hexagonal lattice, with the following detail
amendments. The total number of pixels in the hexagonal support is
(c.f.\ (\ref{sq: N def}))
\begin{equation}
N \equiv 1+3n(n-1) \ .
\end{equation}
Radial index $j$ and circumferential index $i$ transform to
2-component index $\{ij\}$ according to (c.f.\ \ref{2-component
index def})
\begin{equation}
\begin{array}{c}
\{ij\} \equiv i + \min(1,j-1) + 3(j-2)(j-1) \ : \\
j \in [1,n] \ , \ i \in [1,\max(6(j-1),1)] \ , \ \{ij\} \in [1,N]
\ .
\end{array}
\end{equation}
Details of the hexagonal lattice point group---being the dihedral
group $D_{6}$---are presented in
Table~\ref{tab:hex_point_group}\footnote{Derivation of the rank 2
tensor representation of the hexagonal lattice point group
requires the following identity: $(i+j \bmod n) \bmod n = (i+j)
\bmod n$, for $i,j,n=0,1,2,\ldots$, where the definition $\cdot
\bmod 0 \equiv 0$ is adopted.} (c.f.\ Table
\ref{tab:sq_point_group}). Point subgroups of the hexagonal
lattice are: $\{ T_{1}, T_{2}, T_{3}, T_{4}, T_{5}, T_{6}\}$
representing a 6-fold rotation axis; $\{ T_{1}, T_{3}, T_{5},
T_{7}, T_{9}, T_{11}\}$ and $\{ T_{1}, T_{3}, T_{5}, T_{8},
T_{10}, T_{12}\}$ representing 6-fold triangular symmetry; $\{
T_{1}, T_{3}, T_{5} \}$ representing a 3-fold rotation axis; $\{
T_{1}, T_{7}\}$, $\{ T_{1}, T_{8}\}$, $\{ T_{1}, T_{9}\}$, $\{
T_{1}, T_{10}\}$, $\{ T_{1}, T_{11}\}$ and $\{ T_{1}, T_{12}\}$
representing 2-fold reflection symmetry (one mirror line); and $\{
T_{1}\}$ representing complete absence of geometric symmetry.
Depending on the degree of symmetry of the continuous image
ensembles, it may or may not be beneficial to use hexagonal
sampling instead of square sampling. For an hexagonal lattice,
$T^{(2i)}_{1}$ has eigenvalue 1, $T^{(2i)}_{4}$ and
$T^{(2i)}_{7-12}$ have eigenvalues 1 and $-1$, $T^{(2i)}_{3}$ and
$T^{(2i)}_{5}$ have eigenvalues 1 and $-1/2 \pm \sqrt{3}/2 i$, and
$T^{(2i)}_{2}$ and $T^{(2i)}_{6}$ have eigenvalues $\pm 1$, $1/2
\pm \sqrt{3}/2 i$, and $-1/2 \pm \sqrt{3}/2 i$. The only amendment
to the degeneracy matrix analysis of Section~\ref{sec: D matrix}
is that for the hexagonal lattice columns of
$D^{(2)\prime\prime\prime}$ have either 12 1s, or 6 2s, or 1 12
entry; likewise for $D^{(4)\prime\prime\prime}$ with the inclusion
of columns with 3 4 entries. Columns with 4 3s or 2 6s, although
plausible on number theoretical grounds, never occur in
$D^{(2i)\prime\prime\prime}$ (c.f.\ Footnote~\ref{fn: D'''
columns}).

Reduction in detector degrees of freedom consequent on hexagonal
lattice permutation and geometric symmetries is quantified in
Table~\ref{tab:hex_dof} (c.f.\ Table~\ref{tab:sq_dof}). Support
linear dimensions are quantified by $n$ for both square and
hexagonal lattices, but $n$ is akin to the support diameter for
the square lattice and radius for the hexagonal lattice.
Accordingly, square and hexagonal supports with the same $n$
contain very different numbers of pixels---almost 3 times more in
the hexagonal support for large $n$. A hexagonal support of linear
dimension $n$ is closer in pixel population to a square support of
linear dimension $2n$---just over 1/3 more in the square support
for large $n$. Permutation symmetry reduces degrees of freedom by
a factor of almost 2 for larger supports, for both square and
hexagonal lattices\footnote{The reduction in degrees of freedom
due to permutation symmetry is independent of sampling lattice,
but depends on the degree of the detector's polynomial response
function. For an $m$th degree polynomial detector, permutation
symmetry reduces degrees of freedom by a factor of almost $m!$ for
large supports.}. Geometric symmetry reduces degrees of freedom by
a factor of almost $s$ for large supports, for both square and
hexagonal lattices. The hexagonal lattice, with $s\!=\!12$,
achieves greater reduction in degrees of freedom than the square
lattice, with $s\!=\!8$.

\section{Experimental results}
\label{sec: experiment}

The effect of QFD symmetrisation on target detection in real
imagery is investigated. Discrimination is between SAR images of
bushland and SAR images of a variety of vehicles immersed in the
same bushland, samples of which are shown in Figure~\ref{fig:image
chips}. Images are sampled on a square lattice. Image ensembles
are a mixture of different flights of a spotlight SAR illuminating
the ground at different angles of incidence. Clutter images are
not just random samples of the background, but are background
regions that `tricked' a simple prescreener into declaring a
target; hence the difficulty of distinguishing between clutter and
targets in Figure~\ref{fig:image chips}. In these experiments
ensemble sizes are: clutter training---17197 ; clutter
test---37909 ; target training---3425 ; target test---3579.
Detector support is $9\!\times\!9$ pixels, so $n\!=\!9$.
Unsymmetrised and symmetrised QFDs are computed by the same
procedure, but using different degeneracy matrices~$D$. The
unsymmetrised detector degeneracy matrix accounts for only
permutation symmetry, while the symmetrised detector degeneracy
matrix accounts for the full 8-fold square lattice point symmetry
as well as permutation symmetry. It has been noted in
Section~\ref{sec: square lattice} that square lattice sampled SAR
image ensembles strictly have only 2-fold symmetry, so in
principle the detector is being excessively symmetrised here.
Justification for maximally symmetrising the detector is that SAR
image ensembles have approximately 8-fold symmetry, and that
imposing the full 8-fold symmetrisation on the detector allows
verification of the full square lattice point group representation
tabulated in Table~\ref{tab:sq_point_group}. However, there is a
risk that the excessive symmetrisation will reduce the detector
effectiveness compared with its unsymmetrised counterpart.

Experimental `detection error trade-off' (DET)\footnote{DET curves
present the same information as `receiver operating
characteristics' (ROC) curves, but better illustrate low error
regions.} curves are plotted in Figure~\ref{fig:exp_res} in
various combinations to assist comparison. Error bars ($\pm1$
standard deviation) are plotted in graphs that compare training
and test ensemble DET curves (i.e.\ graphs (a) and (b)). Graph (a)
demonstrates that the unsymmetrised QFD generalises poorly, while
graph (b) demonstrates that the symmetrised QFD generalises well.
The reason for this behaviour is revealed in
Table~\ref{tab:sq_dof}, which states that the unsymmetrised QFD
has 3402 degrees of freedom compared with the 476 of the
symmetrised QFD\@. There is more scope for the unsymmetrised QFD
to overtrain on the training ensemble than there is for the
symmetrised QFD, as is apparent in graph (c). That the
unsymmetrised QFD is going to generalise poorly in this example,
is predictable beforehand on the basis of the number of detector
degrees of freedom (3402) not being much less than the size of the
smaller of the target and clutter training ensembles (3425). Both
unsymmetrised and symmetrised detectors should have their
generalisation improved by regularisation
procedures~\cite{Friedman89, Hastie95}. Graph (d) demonstrates
that---despite the caution expressed at the end of the previous
paragraph---the symmetrised QFD is more effective than the
unsymmetrised QFD when tested on the test ensembles, and the
advantage is most pronounced at the important low false-alarm
probabilty extreme of the DET\@. The correctness of the square
lattice point group representation tabulated in
Table~\ref{tab:sq_point_group} has been verified, and the
practical value of QFD symmetrisation---even theoretically
excessive symmetrisation---has been established.

\section*{Acknowledgements}

The author is grateful to Warwick Holen for drafting Figures
\ref{fig: sq_net} and~\ref{fig: hex_net}, Nick Stacy for
furnishing the DSTO Ingara SAR images used in the experiment of
Section~\ref{sec: experiment}, and Martin Oxenham for constructive
criticism of a preliminary draft of this article.

\vspace*{-2\baselineskip}

\begin{biographynophoto}
{Robert Caprari} was awarded a Bachelor of Engineering (Honours)
degree in electrical and electronic engineering, and a Bachelor of
Science degree, from Adelaide University. He was awarded a Doctor
of Philosophy degree in physics from Flinders University, where
his doctoral research was in electron scattering and condensed
matter physics. All of his professional career has been spent as a
research scientist with the Defence Science and Technology
Organisation (DSTO) in Australia. Within DSTO, his research has
broadly been in the fields of signal theory, image processing,
pattern recognition and statistical learning.
\end{biographynophoto}

\vfill

\end{document}